# A Geometric Interpretation of the Boolean Gilbert-Johnson-Keerthi Algorithm

Jeff Linahan

**Abstract**

The Gilbert-Johnson-Keerthi (GJK) algorithm is an iterative improvement technique for finding the minimum distance between two convex objects. It can easily be extended to work with concave objects and return the pair of closest points. [4] The key operation of GJK is testing whether a Voronoi region of a simplex contains the origin or not. In this paper we show that, in the context where one is interested only in the Boolean value of whether two convex objects intersect, and not in the actual distance between them, the number of test cases in GJK can be significantly reduced. This results in a simpler and more efficient algorithm that can be used in many computational geometry applications.

## 1. Introduction

We start with a few essential definitions that will help us understand the outline of the Gilbert-Johnson-Keerthi algorithm and our proposed modifications. Throughout this paper we will refer to the Gilbert-Johnson-Keerthi algorithm as GJK.

## 1  Configuration Space Obstacle (CSO)

An *object* is a compact, non-empty set of points in three dimensional Euclidean space. Note that an object is not necessarily polyhedral; it may be curved. The *Minkowski sum* A+B of two objects $A$ and $B$ is determined by adding every point in $A$ to every point in $B$: A+B = {$x$+y: $x \in A, y \in B$}. The Minkowski sum can be visualized geometrically by sweeping one object around the other:

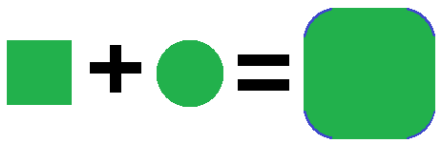

For an object $A$, we use the notation $-A$ to denote the reflection of $A$ about the origin $O$. The *Minkowski difference* A-B can be obtained by computing the Minkowski sum of $A$ and -$B$. An important property used by the GJK algorithm is captured by the following lemmas:

**Lemma 1.** Assume $A$ and $B$ are two convex objects. The Minkowski sum A+B (and Minkowski difference A-B) is also a convex object.

**Lemma 2.** The Minkowski difference of two objects contains the origin if and only if the two objects intersect.

Lemma 1 has been established in [5]. Lemma 2 follows immediately from the fact that, if and only if A and B intersect, there exists a point common to A and B. Lemma 2 enables GJK to operate on the Minkowski difference between two objects, rather than on the objects themselves. The Minkowski difference is also called the *configuration space obstacle* (CSO) [6]. Thus the goal of GJK reduces to determining whether the CSO contains the origin or not.

The beauty of GJK is that it does not actually calculate the CSO. Instead, GJK only samples the CSO by iteratively modifying a *simplex* inside the CSO until the simplex either encloses the origin or the algorithm proves the origin cannot be enclosed. For a fixed integer $k \geq 0$, a $k$-simplex is defined as the convex hull of $k+1$ points in $k$-dimensional space. Thus a point is a 0-simplex, a line segment is a 1-simplex, a triangle is a 2-simplex and a tetrahedron is a 3-simplex. Carathéodory's theorem shows that a simplex inside the object which encloses the origin can always be found so long as the object actually contains the origin, and thus that the algorithm terminates.

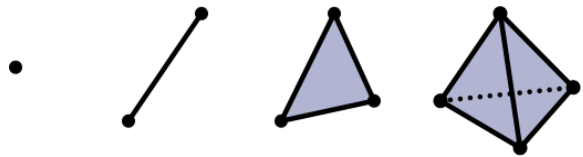

For our purposes, $k = 2$ and $k = 3$ are the natural choices corresponding to two-dimensional and three-dimensional spaces. The construction of a simplex makes use of a *support function*, which we define next. A plane divides the space into two *half-spaces*. We say that a plane $P$ *supports* an object $A$ if the following two conditions hold:
- $A$ is entirely contained in one of the two closed half-spaces determined by $P$
- At least one boundary point of $A$ lies in $P$

Note that supporting planes are tied to a specified direction, so it is natural to define $S_A(\mathbf{v})$ associated with an object $A$ mapping a direction vector $\mathbf{v}$ to the signed distance the supporting plane is from the origin:

$$S_A(v) = \max\{v \cdot a : a \in A\}$$

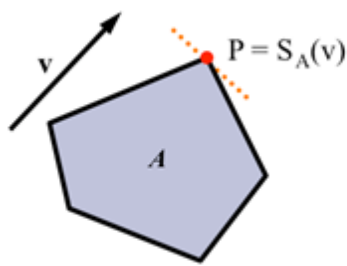

Support functions satisfy two important properties:

**Lemma 3:** Let $C$ be the Minkowski sum of convex objects $A$ and $B$. The support of $C$ is the sum of the supports of $A$ and $B$: $S_{A+B}(\mathbf{v}) = S_A(\mathbf{v}) + S_B(\mathbf{v})$ [5].

**Lemma 4:** Let *D* be the Minkowski difference of two convex objects *A* and *B*. The support of *D* is the support of A minus the support of B with an opposite direction vector: $S_{A-B}(\mathbf{v}) = S_A(\mathbf{v}) - S_B(-\mathbf{v})$.

Lemma 3 follows from the definition of support function. Lemma 4 follows immediately from the observation that $S_{-B}(\mathbf{v}) = -S_B(-\mathbf{v})$. Recall that the input to GJK is not two objects, but the CSO of the two objects. Of course, the CSO contains an infinite number of points (unless the objects intersect at exactly one point or don't intersect at all) so the input must be given in the form of the CSO's support function, which, by Lemma 4, is defined as:

$$S_{CSO}(\mathbf{v}) = S_A(\mathbf{v}) - S_B(-\mathbf{v})$$

Note that while support functions are usually defined in the literature as the supporting plane's distance to the origin, GJK requires we know the point on the plane. A *support point* of an object *A* is a point on *A*'s surface that is farthest along a given direction. Support points are critical to constructing the GJK simplices, so hereafter support functions will be assumed to return the point on the supporting plane. This construction poses one problem though: for polyhedral objects support points are not necessarily unique. If the direction vector is parallel to a face normal, every point on that face is equally extremal in that direction, and hence, is a support point. In this case we simply require the support function return the centroid of the face in question. As we shall shortly see, the GJK algorithm operates only on certain points on the CSO's surface, which are computed on demand.

## 2   Overview of the Original GJK Algorithm

The original GJK algorithm [8] described by Gilbert, Johnson and Keerthi was restricted to computing the minimum distance between two convex polytopes. It could easily be extended to work with concave objects constructed with unions of convex objects. Later, Gilbert and Foo [9] enhanced the original algorithm to handle general curved convex objects. In this section we provide a brief description of the enhanced GJK algorithm by Gilbert and Foo [9] as it appears in [6]. The key operation in this original presentation is finding an object's *point of minimum norm*, the object's point closest to the origin. Computing the distance between two convex objects reduces to computing the distance from their CSO to the origin. (see Lemma 2). Therefore, we must find the point of minimum norm of the CSO, which we will refer to as `MinimumNormPoint(CSO)`. We use the standard notation $\|P\|$ to denote the *norm* (distance to the origin) of a point P. The GJK algorithm iteratively constructs simplexes inside the CSO closer to the origin until the simplex ceases to change. The point P of minimum norm of these simplexes serves as successive approximations to `MinimumNormPoint(CSO)`. For polytopes, GJK terminates after a finite number of iterations. For curved objects, an epsilon term must be added to prevent infinite loops and put a bound on the error between the computed and actual separating distance. It has been established in [6] that the squared distance between the approximation P and actual `MinimumNormPoint(CSO)` is bounded to be below `P·V`, where V is the support point of the CSO in the direction –P. Thus if $\|P\|^2 - P \cdot V \leq \varepsilon^2$, for some fixed very small real constant $\varepsilon \geq 0$, we conclude that V is no more extreme in the direction –P than P itself and the distance from the CSO to the origin is $\|P\|$. The GJK algorithm is sketched below:

```
1. P ← any point in the CSO
2. Q ← {}
3. V ← S_CSO(-P)

4. while ||P||² - P·V > ε²
       P ← MinimumNormPoint(ConvexHull(Q ∪ {V}))
       Q ← smallest Q' ⊆ Q ∪ {V} such that P ∈ ConvexHull(Q')
       V ← S_CSO(-P)
5. return ||P||
```

The GJK algorithm starts with an arbitrary point P in the CSO and empty simplex set Q, then computes a support point V in the direction −P. If V is no further in the direction −P than P itself, then P must be the point closest to the origin: the algorithm terminates with ‖P‖ as the distance from the CSO to the origin. Otherwise, the algorithms adds V to the current simplex set Q and updates P to be the point closest to the origin within the new simplex (`MinimumNormPoint(ConvexHull(Q))`). The simplex Q is reduced to the smallest convex subset of Q still containing P, and the support point V in direction −P is updated for the next iteration.

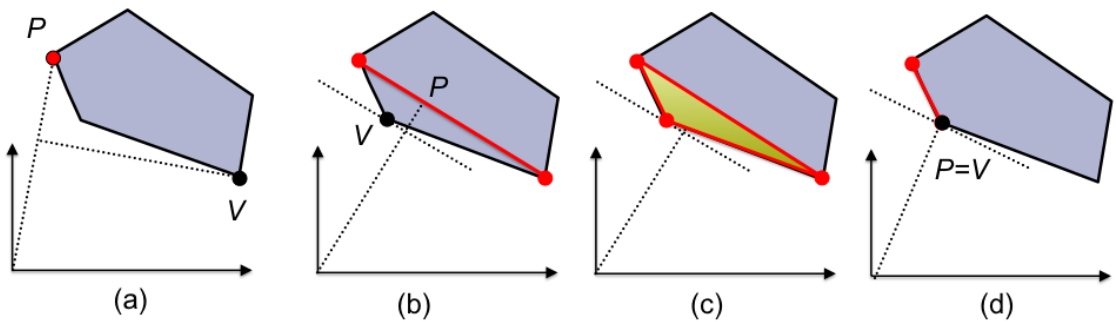

The figure above depicts the steps of the algorithm on a very simple example. The algorithm starts with the values of P and V marked in (a), and the first iteration updates the simplex to the line segment connecting these two points. In the second iteration, the point of that simplex closest to the origin is updated to P in (b), and the support point in direction −P is V. The simplex grows to the triangle formed by the tree points shaded in (c), but this can be reduced to the line segment containing P highlighted in (d). In this third and final iteration, the algorithm finds no point closer than P in direction −P and terminates.

One area requiring further elaboration is how to compute the point of minimum norm. This is accomplished with *Johnson's distance subalgorithm* [10], a mathematical optimization technique that works by solving systems of linear equations. Unfortunately, Johnson's distance subalgorithm suffers from numerical robustness problems.

## 3   Simplified GJK to Test Intersections

Ericson [4] observed that the GJK algorithm could be simplified and made more robust by substituting a geometric distance subalgorithm for Johnson's algebraic one. His version finds the point of minimum norm using primitive geometric functions such as

`ClosestPointOnEdgeToPoint()` and `ClosestPointOnTriangleToPoint()`, which are implemented with vector operations such as dot products. Other than that, his algorithm version is very similar to the original GJK version, with the exception of a small optimization: an early termination condition testing whether the point P closest to the origin coincides with the origin: in this case, the objects must intersect (see Lemma 2) and the algorithm returns a distance of zero.

## 4  Boolean GJK

Muratori [2] presented a version of GJK for the situation where the application only needs to test whether two objects intersect or not, and the separation distance is irrelevant to the application. For example, GJK may be used as a broad phase collision detection technique on convex hulls of polyhedral objects to reduce the number of object pairs needing an exhaustive triangle vs. triangle test. As another example, a computer graphics application may use GJK to perform view frustum culling. In these cases, the algorithm returns a Boolean value (true if the objects intersect, false otherwise). We will refer to this algorithm version as *BGJK* (*Boolean GJK algorithm*).

```
1. Initialize support point S ← S_CSO(any direction)
2. Initialize simplex ordered set Q ← {S}
3. Initialize direction D ← -S
4. Loop:
     S ← S_CSO(D)
     If S·D < 0 then return false
     Q ← Q ∪{S}
     If DoSimplex(Q, D) then return true
```

Similar to regular GJK, we begin by initializing a simplex with a point S generated using an arbitrary search direction. S is then used to generate the new direction D from S to the origin. The main loop consists of three main steps. First, a new support point S is computed based on the current search direction D. Next we compare this new point S with a plane passing through the origin and perpendicular to D. If S is not in the closed half-space that D is pointing into (`S·D < 0`), then we conclude that the CSO has no points on D's side of the plane (or the plane itself.) This is because we looked for a support point that was as far into D's half-space as possible, but could not even get to the separating plane. Therefore the CSO does not contain the origin and the objects do not intersect. Otherwise, if S is in D's closed half-space, we add it to the simplex set Q can continue trying to enclose the origin. The last step is the `DoSimplex` procedure, Muratori's major contribution. It brings up a few important insights into the Boolean version of the GJK algorithm. The first key observation is that the point of minimum norm *P* need not be computed explicitly. To accomplish this, `DoSimplex` has the following responsibilities:

1. Checks which Voronoi region of Q contains the origin. If Q itself contains the origin, then the objects must be intersecting (by Lemma 2) and the function returns true. Note the difference from Ericson's algorithm version, which tested whether P *coincided* with the origin.

2. Removes any points from the simplex Q that are now unnecessary. This includes all points that do not make up the subsimplex that whose Voronoi region contains the origin.

3. Chooses a new direction D = -P to expand the simplex in on the next iteration. The method to compute D depends on the size of the new subsimplex (here we are concerned with the 0-simplex, 1-simplex, and 2-simplex). If P is a subsimplex vertex (i.e., the origin is in P's Voronoi region) then D is a vector from vertex P to the origin. Otherwise, P is on another subsimplex feature (a line segment or a triangle.)  In this case, a vector perpendicular to that feature and pointed towards the origin (generated using dot and cross products) is equivalent to -P.

The second key observation Muratori reported is that it is not necessary to test all Voronoi regions of the simplex in each iteration. In particular, if one implements GJK in such a way that it remembers the order in which vertices were added to the simplex (instead of simply considering it a set), one can prove that the origin cannot possibly be in several Voronoi regions that open backwards with respect to the search direction, and omit those tests.

## 2. Our Contribution

A third key insight due to Rory Driscoll further reduces the number of Voronoi regions used in Muratori's BGJK. Here we will prove that Muratori and Driscoll's optimizations are valid, leading to a significantly simpler and more efficient implementation. We will follow the convention of naming the simplex points alphabetically, with the most recently added point as A. We will now examine the behavior of `DoSimplex` when called with n-simplices for various n, where n+1 is the cardinality of the ordered simplex set Q.

### 1-Simplex

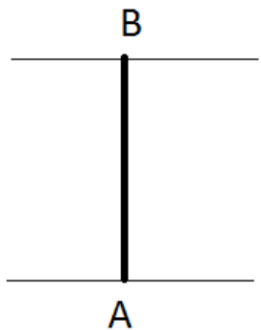

**Top View**

In the case of 2 points, Q = {A, B} is a 1-simplex (line segment) with 3 Voronoi regions: two half-spaces and a slab.

**Lemma 5:** If the CSO contains the origin when n = 1, the slab Voronoi region must contain the origin.

*Proof.* Suppose we hit a breakpoint upon entering the `DoSimplex` subroutine when n=1. Prior to entering the loop, D was generated by reflecting the initial support point (now called B) about the origin, so if this is the first time `DoSimplex` was called we have D = -B. If this is not the first time, we know the last time `DoSimplex` was called it set Q to a 0-simplex {B} because it is currently a 1-simplex {A, B}, with S = A being the most recently added point. In this case, recall that `DoSimplex` is required to set the search direction D to -P, which in the 0-simplex case is equal to -B, so D always equals -B at this breakpoint when n=1. We also know that the dot product termination conditional failed, so S·D ≥ 0. By substitution we have A· -B ≥ 0, or A · B ≤ 0. By Thales' Theorem the region of R3 where A · B ≤ 0 is the sphere with line segment AB as a diameter, a subset of the slab Voronoi region of line segment AB. ∎

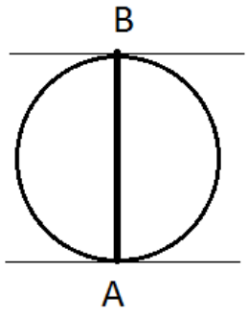

**Top View**

  `DoSimplex` constructs a new search direction that points from line segment AB to the origin by orthogonalizing -A with respect to AB:

$$D = (AB \times -A) \times AB$$

`DoSimplex` leaves the 1-simplex unchanged in this case since AB is always the closest feature to the origin.

**2-Simplex**

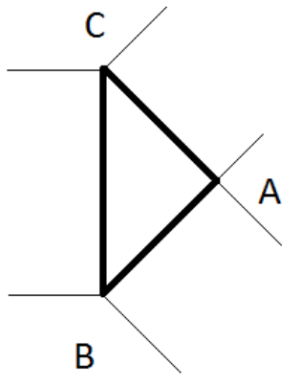 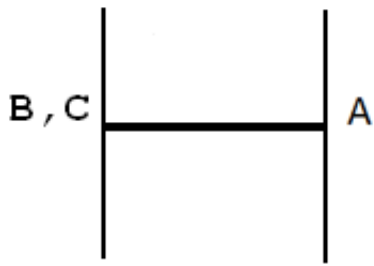

**Top View**        **Side View**

In the case of 3 points, Q = {A, B, C} is a 2-simplex (triangle) with 8 Voronoi regions: one for each of the three vertices, one for each of the three edges, and two for the regions above and below the triangle.

**Lemma 6:** If the CSO contains the origin when n = 2, either the AC, AB, ABC, or ACB Voronoi regions must contain the origin.

*Proof.* Suppose we hit a breakpoint upon executing DoSimplex when n=2. From the 1-simplex case, we know that the origin must be in the sphere with diameter BC, so it cannot be in region B or C. We also know that D·A ≥ 0, (D points to the right with respect to the figures), because otherwise S(D) would not have generated A this iteration. (B and C are farther left, for example.) Therefore the origin must be in the hemisphere with diameter BC and plane of symmetry ABC. This implies the origin is not in region BC. Furthermore, if the origin was in region A then A·D ≤ 0 and the termination condition would have exited the algorithm already this iteration. This leaves only AC, AB, ABC, and ACB. ∎

If the closest feature is an edge, the point opposite that edge is deleted from the simplex and the search direction is orthogonalized with respect to this edge, pointed towards origin. If the origin lies in one of two the triangle's Voronoi regions, either the simplex is unchanged, or the winding of the vertices is swapped because it is significant whether it is above or below the triangle. The search direction is set to the triangle's normal, then multiplied by -1 if the vertex winding was swapped.

**3-Simplex**

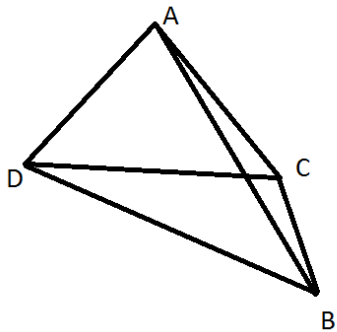 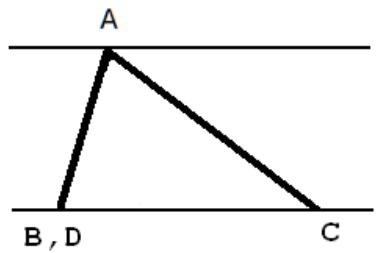

**Perspective View**          **Side View**

In the case of 4 points, Q = {A, B, C, D} is a 3-simplex (tetrahedron) with 15 Voronoi regions: one for each of the four vertices, one for each of the six edges, one for each of the four triangles, and the internal region.

**Lemma 7:** If the CSO contains the origin when n = 3, either one of the Voronoi regions of the 3 faces and 3 edges that contain A or tetrahedron must contain the origin.

*Proof.* Suppose we hit a breakpoint upon executing DoSimplex when n=3. As with the 1- and 2-simplex cases, the origin must be in the hemisphere with diameter DB and axis of symmetry lying in plane BCD, so it cannot be in regions D, B, or DB. Our last proof also showed it cannot be in region C due to the dot product termination condition. Using similar reasoning it cannot be in region A; S(D) = A this iteration, and we know D must be the normal of triangle BCD. With respect to the side view, the search direction D is pointed up, so the origin cannot be in the Voronoi region below triangle CBD, otherwise we would have swapped the triangle's winding and turned our frame of reference upside down. Lastly, if the origin were in Voronoi region BC, then the previous iteration of `DoSimplex` would have discarded point D from the simplex and we would not be in the 3-simplex case. Likewise, if the origin were in Voronoi region DC, then point B would have been discarded. This leaves Voronoi regions ABCD, ABC, ACD, ADB, AB, AC, and AD. ∎

The code determining in which of the 7 Voronoi regions the origin lies requires 6 if statements, but fortunately we can engineer these tests in a divide and conquer fashion such that at most only 3 statements are tested each time `DoSimplex` is invoked. This amounts to a simple plane vs. point test on each of the tetrahedron's sides except CBD.

A noteworthy optimization that our proofs imply is that the n = 1 `DoSimplex` case can be unrolled from the loop because once the algorithm reaches the n = 2 case, it will never return to the case where n = 1, illustrated below:

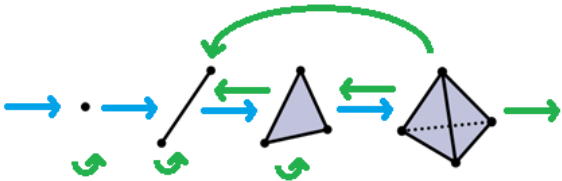

Arrows pointing right (blue) represent adding S to Q, all others (green) represent `DoSimplex` optionally removing vertices from Q (or returning true in the 3-simplex case.) Each iteration consists of taking a blue step and then a green step. Note how after we reach the 2-simplex (triangle), we never return to the 1-simplex (line segment) again.

**Conclusion**

We have shown that the number cases in implementing the Gilbert-Johnson-Keerthi algorithm can be significantly reduced in the case where one is interested only in a Boolean result of whether two objects in 3D space are intersecting. As computer graphics applications become more sophisticated, we find that studying special cases of algorithms for special cases in three dimensional space leads to important insights on how to efficiently and robustly implement them.